# Coexistence of topological surface states and superconductivity in Dirac semimetal NiTe₂

Chen He, [1,§,*] Jian-Zhou Zhao,[2,*] Mei Du,[1,*] Luo-Zhao Zhang,[3] Jia-Ying Zhang,[3] Kuo Yang,[1] Noah F. Q. Yuan,[1] Aleksandr Seliverstov,[4] Ewald Janssens,[4] Jun-Yi Ge,[3,†] Zhe Li[1,5,‡]

[1]State Key Laboratory on Tunable laser Technology, Ministry of Industry and Information Technology Key Lab of Micro-Nano Optoelectronic Information System, School of Science, Harbin Institute of Technology (Shenzhen), Shenzhen, 518055, China
[2]Co-Innovation Center for New Energetic Materials, Southwest University of Science and Technology, Mianyang 621010, China
[3]Materials Genome Institute, Shanghai University, 200444 Shanghai, China
[4]Quantum Solid-State Physics, Department of Physics and Astronomy, KU Leuven, Celestijnenlaan 200D, 3001, Leuven, Belgium
[5]Guangdong Provincial Key Laboratory of Semiconductor Optoelectronic Materials and Intelligent Photonic Systems, Harbin Institute of Technology (Shenzhen), Shenzhen 518055, China
*These authors contributed equally.
§Present Address: Quantum Solid-State Physics, Department of Physics and Astronomy, KU Leuven, BE-3001 Leuven, Belgium
†Corresponding authors: junyi_ge@t.shu.edu.cn
‡Corresponding authors: zhe.li@hit.edu.cn

†Contact author: junyi_ge@t.shu.edu.cn
‡Contact author: zhe.li@hit.edu.cn

**ABSTRACT**. The coexistence of topological bands around the Fermi level ($E_F$) and superconductivity provides a fundamental platform for exploring their interplay. However, few materials inherently display both properties. In this study, we demonstrate the coexistence of topological surface states at the $E_F$ and superconductivity in $NiTe_2$ single crystals, a material hitherto not recognized as superconducting. Quasiparticle interference measurements performed via scanning tunneling microscopy suggest the presence of topological surface states at the $E_F$, which is further corroborated by density functional theory simulations. Experimental evidence for superconductivity is provided via electronic transport measurements and specific heat capacity analyses. Our results suggest that $NiTe_2$ represents a promising platform for investigating the rich interplay between topological states and superconductivity.

Topological states in materials are protected by the topological invariants of their band structures and exhibit remarkable robustness against local perturbations, defects, or impurities [1,2]. The introduction of superconductivity into these topological states, particularly when they reside near the Fermi level ($E_F$), offers a promising route toward realizing topological superconductivity [3]. However, materials that exhibit both topological states residing near the $E_F$ and superconductivity are rare, especially among stoichiometric compounds.

Recently, $NiTe_2$ crystals have been reported as type-II Dirac semimetals, hosting a bulk type-II Dirac cone [4-7], topological surface states crossing the $E_F$ [4-7], and possible topological hinge states [8]. In $NiTe_2$ based Josephson junctions (JJs), superconducting currents injected from superconducting electrodes into $NiTe_2$ couple with topological surface and hinge states, exhibiting possible evidence of finite momentum Cooper pairing [9] and hinge supercurrents [8]. These studies suggest that $NiTe_2$ may serve as a promising platform for investigating the interplay between topological states and superconductivity. Although superconductivity in bulk $NiTe_2$ was reported to be induced via chemical doping [10] or applied pressure in $NiTe_{2-x}$ ($x = 0.38 \pm 0.09$) [11], pristine stoichiometric $NiTe_2$ have not been recognized as superconducting [8,9,12,13].

In this work, we demonstrate that stoichiometric $NiTe_2$ single crystals exhibit both topological surface states at $E_F$ and intrinsic superconductivity. The topological surface states are characterized using scanning tunneling microscopy (STM), supported by density functional theory (DFT) calculations, while intrinsic superconducting properties are identified via electronic transport measurements and specific heat capacity analyses. The coexistence of topological surface states at the $E_F$ and superconductivity in stoichiometric $NiTe_2$ provides new opportunities to explore the interplay between topology and superconductivity in van der Waals materials.

$NiTe_2$ single crystals [see Fig. S1 in the Supplemental Material (SM)] were synthesized with average atomic concentrations of 33.7% Ni and 66.3% Te, which closely approximate the stoichiometric ratio. The surface of $NiTe_2$ was characterized using an STM operating at 4.5 K under ultrahigh vacuum conditions. Prior to characterization, the $NiTe_2$ single crystal was cleaved *in situ* along the (001) plane, thereby exposing the topmost Te atoms. Figures 1(a) and 1(b) display the large-area and atomic-resolution topographies of the cleaved $NiTe_2$ surface, respectively. The inset in Fig. 1(b) shows the fast-Fourier transform (FFT) of the atomic-resolution image. From the FFT image, we extracted a lattice constant for the top layer of $NiTe_2$ of a = 3.99 ± 0.3 Å, which is in close agreement with the theoretical value of 3.86 Å and previous study [14]. The white hexagon in the inset of Fig. 1(b) delineates the surface Brillouin zone (SBZ). The differential conductance ($dI/dV$) spectroscopy in Fig. 1(c) reflects the electronic density of states (DOS) on the surface, revealing a general suppression of intensity around the $E_F$. Additionally, three peaks ($\alpha$, $\beta$, and $\gamma$) are observed near $E_F$, which is consistent with previous reports [15]. Based on the calculated spin-polarized band structure presented in Figs. 1(d)–(f), the $\alpha$, $\beta$, and $\gamma$ peaks can be attributed to enhanced contributions from topological surface states. These features in the $dI/dV$ spectroscopy are well reproduced by the DFT-calculated DOS [red curve in Fig. 1(c)].

To elucidate the spin texture of the topological surface states in $NiTe_2$, we present the spin-resolved band structure projected onto the topmost Te atom layer, as shown in Figs. 1(d)–(f). Our observations reveal that the Fermi surface is primarily composed of two distinct topological surface states, designated as TSS1 and TSS2. Both states originate from band inversions occurring above and below the $E_F$ [5]. TSS1 is characterized by the presence of both electron and hole pockets (*e*–pocket and *h*–pocket). In contrast, the $S_z$ spin component of TSS2, as shown in Fig. 1(f), indicates that its two intersecting linear dispersions exhibit opposite spin-momentum locking, leading to the formation of a tilted, Dirac-cone-like feature at $E_F$. Furthermore, the $S_z$ projection along the $\overline{\Gamma} - \overline{M}$ direction shows no band dispersion, suggesting that the spins of both TSS1 and TSS2 are pinned to the in-plane direction.

†Contact author: junyi_ge@t.shu.edu.cn
‡Contact author: zhe.li@hit.edu.cn

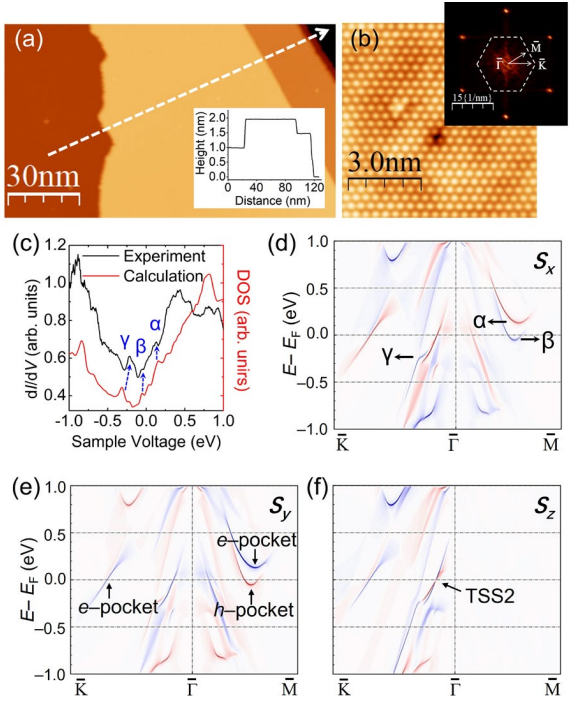

FIG. 1. Topography and band structure of the NiTe$_2$ surface. (a) A 138 nm × 81 nm STM topography image of the in situ cleaved NiTe$_2$ surface (set point: V = 1 V, I = 0.02 nA). The inset displays the height profile taken along the indicated dashed arrow. (b) Atomic-resolution STM topography image showing the topmost Te atoms on the NiTe$_2$ surface. The inset presents the fast-Fourier transform (FFT) map of (b), highlighting the surface Brillouin zone (SBZ) and the identified $\overline{\Gamma} - \overline{K}$ and $\overline{\Gamma} - \overline{M}$ directions. (c) d$I$/d$V$ spectroscopy (black curve) and the calculated density of states (DOS) (red curve). (d)-(f) Spin-resolved band structure projected onto the topmost Te atom layer along the $S_x$, $S_y$, $S_z$ direction, respectively. The $S_x$, $S_y$, $S_z$ directions correspond to the Cartesian coordinate system defined for the NiTe$_2$ lattice in real space ($S_x$: along the [110] direction; $S_y$: along the [100] direction (in-plane); $S_z$: along the [001] direction).

To explore the detailed features of these topological surface states near $E_F$, we conducted STM-based quasiparticle interference (QPI) measurements. Figures 2(a)–(c) display the experimental QPI patterns at 0 meV, +103 meV, and +198 meV, respectively. The corresponding real-space d$I$/d$V$ grid can be found in Fig. S2 of the SM. Due to resolution limitations, the scattering vector near the $\overline{\Gamma}$ point cannot be adequately resolved in Figs. 2(a)–2(c). The primary scattering vectors identified in the QPI patterns are $q_1$, $q_1'$, and $q_2$, which correspond to scattering processes involving relatively large momentum changes [Figs. 2(a)–2(c)].

†Contact author: junyi_ge@t.shu.edu.cn
‡Contact author: zhe.li@hit.edu.cn

As the energy increases from 0 meV to +198 meV, the scattering vector $q_1$ gradually shifts from the $\overline{\Gamma} - \overline{K}$ direction to the $\overline{\Gamma} - \overline{M}$ direction, eventually evolving into $q_1'$. Another notable feature is that, with increasing energy, the intensity of the scattering vector $q_2$ gradually weakens and eventually disappears.

To understand the origin of the scattering vectors $q_1$, $q_1'$, and $q_2$ observed in the QPI measurements, we present the calculated spin-resolved constant energy surface maps at these energy levels in Figs. 2(d)–(i). In the constant energy surface maps at 0 meV [Figs. 2(d), 2(g), and 2(j)], TSS1 is identified by the "w"-shaped $e$–pockets near the $\overline{\Gamma} - \overline{K}$ direction and $h$–pockets around the $\overline{\Gamma} - \overline{M}$ direction, whereas TSS2 is characterized by a closed hexagon near the $\overline{\Gamma}$ point. These spin-resolved constant energy surface maps provide detailed information on the spin texture of the topological surface states. For the in-plane spin components [Figs. 2(d) and (g)], the spins associated with both TSS1 (in the $e$– and $h$–pockets) and TSS2 rotate around the $\overline{\Gamma}$ point, completing one full winding as the momentum vector $k$ traverses the Brillouin zone. The in-plane spin of the $e$–pocket and that of TSS2 rotate clockwise, whereas the spin of the $h$–pocket rotates counterclockwise. The out-of-plane spin component is observed only in the electron pocket and TSS2 [Fig. 2(m)]. This out-of-plane spin component exhibits threefold symmetry in $k$-space. As the momentum vector $k$ completes a full 360° loop around the $\overline{\Gamma}$ point in the first Brillouin zone, the out-of-plane spin undergoes three full windings. Furthermore, the $h$–pocket and TSS2 exhibit the same out-of-plane spin winding direction.

By comparing the QPI pattern and the spin-resolved constant energy surface, we infer that $q_1$ and $q_2$ likely originate from scattering between the $e$–pockets and $h$–pockets, respectively. The evolution of the scattering vectors, along with the spin-resolved constant energy surfaces, further corroborates our conjecture. With increasing energy, the state density of the $e$–pocket gradually shifts toward the $\overline{\Gamma} - \overline{M}$ direction, leading to a weakening of $q_1$ and the emergence of $q_1'$ at approximately +100 meV, with $q_1$ completely evolving into $q_1'$ at around +200 meV. As for the $h$–pockets, they gradually disappear as the energy increases, causing the $q_2$ scattering vector to gradually vanish. Regarding TSS2, due to its proximity to the $\overline{\Gamma}$ point, accurate QPI scattering information could not be extracted. The combination of the experimental QPI and the calculated spin-resolved constant energy surfaces suggests the existence of topological surface states, i.e. $e$–pocket and $h$–pocket, at the $E_F$.

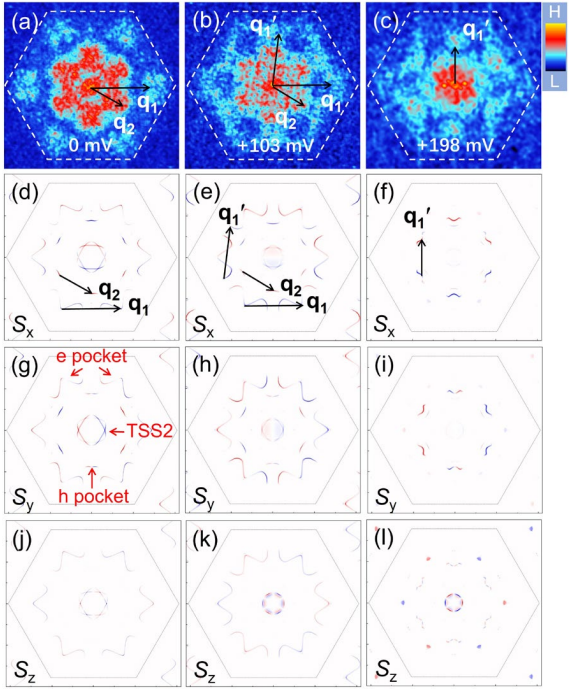

FIG. 2. Quasiparticle interference (QPI) pattern of a NiTe$_2$ surface. (a)–(c) Fast Fourier transform (FFT) d$I$/d$V$ maps at 0 meV, +103 meV, and +198 meV, respectively. Set point: $V = 500$ mV, $I = 2$ nA. The white dashed hexagon indicates the SBZ. (d)–(f) Simulated spin-resolved constant energy surface maps ($S_x$ component) at 0 meV, +100 meV, and +200 meV, respectively. (g)–(i) Simulated spin-resolved constant energy surface maps ($S_y$ component) at 0 meV, +100 meV, and +200 meV, respectively. (j)–(l) Simulated spin-resolved constant energy surface maps ($S_z$ component) at 0 meV, +100 meV, and +200 meV, respectively. The black hexagons indicate the SBZ. The black and red arrows indicate the scattering vectors and topological surface states, respectively.

In addition to the topological surface states, we also uncover the presence of intrinsic superconductivity in NiTe$_2$. As shown in Fig. 3(a), the temperature-dependent resistivity, $\rho(T)$, of NiTe$_2$ displays metallic behavior from 300 K down to 300 mK; however, superconductivity emerges below 300 mK. In Fig. 3(b), $\rho(T)$ below 400 mK is plotted under varying magnetic fields applied along the $c$ axis ($H_{\parallel c}$). The superconducting transition is evidenced by a sharp drop in resistivity approaching the zero-resistance state. We define the critical temperature ($T_c$) and the upper critical field ($H_{c2}$) as the conditions at which the resistivity reaches half of its value at $T = 400$ mK.

†Contact author: junyi_ge@t.shu.edu.cn
‡Contact author: zhe.li@hit.edu.cn

Under zero magnetic field, superconductivity emerges below $T_c = 261$ mK, progressively shifting to lower temperatures and eventually being entirely suppressed with increasing $H_{\parallel c}$. Figure 3(c) shows the field-dependent resistivity, $\rho(H)$, at different temperatures. At 50 mK, $H_{c2}$ is approximately 28 Oe; as the temperature increases, $H_{c2}$ shifts to lower field values. The phase diagram presented in Fig. 3(d) is extracted from both the $\rho(T)$ curves [Fig. 3(b)] and the $\rho(H)$ curves [Fig. 3(c)]. The Werthamer–Helfand–Hohenberg (WHH) model was employed for both single-band and two-band fitting of $H_{c2}(T)$ [16] (fitting details can be found in the Supplementary Information). Our analysis reveals that $H_{c2}(T)$ follows a two-band WHH model [red curve in Fig. 3(d)] rather than a single-band model [blue curve in Fig. 3(d)], suggesting that NiTe$_2$ may host two bulk superconducting bands. The fitting yields a band-coupling parameter $\omega = 24.3$ and a zero-temperature upper critical field $H_{c2}(0) = 34$ Oe. The positive value of $\omega$ indicates the two superconducting bands might be dominated by intraband coupling [17]. Furthermore, $H_{c2}(0)$ is well below the Pauli limit ($H_p(0) = 1.86T_c = 4836$ Oe), indicating that the Cooper pairs are rapidly disrupted by the magnetic field through the orbital pair-breaking mechanism, likely due to the small effective electron mass in NiTe$_2$ [4,18].

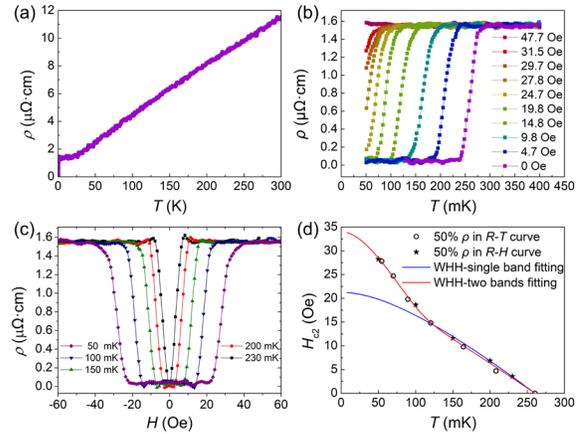

FIG. 3. Temperature-dependent resistivity, $\rho(T)$, and field-dependent resistivity, $\rho(H)$, demonstrating superconductivity in NiTe$_2$. (a) $\rho(T)$ measured from 300 K to 50 mK. (b) $\rho(T)$ below 400 mK under different magnetic field strengths applied along the $c$ axis ($H_{\parallel c}$). (c) $\rho(H)$ in the superconducting phase at various temperatures. (d) Phase diagram, $H_{c2}(T)$, extracted from $\rho(T)$ and $\rho(H)$ measurements.

To further characterize the superconductivity in NiTe$_2$, we measured the specific heat capacity, $C_p$. We present the temperature-dependent specific heat, $C_p(T)$, measured from 150 K down to 70 mK [Figs. 4(a) and 4(b)]. The only distinct feature observed in $C_p(T)$ is a jump around 220 mK. When a $H_{\parallel c}$ is applied, the jump is suppressed to lower temperatures and eventually disappears, which is a characteristic signature of the superconducting transition. To get the details of the phase change, we plot $C_p(T)/T$ in Fig. 4(c). Notably, at $H_{\parallel c}$ = 100 Oe, where the superconducting transition should be fully suppressed [Figs. 3(b)–3(d)], we still observe a divergence in $C_p(T)/T$ toward low temperatures below 700 mK. This feature is further affected by $H_{\parallel c}$, shifting to higher temperatures and exhibiting a series of peaks [Fig. S3 in SM]. Such anomaly in $C_p(T)/T$ is typically attributed to a nuclear Schottky anomaly [19], spin fluctuations [20], or the excitation of magnetic defects [21]. Considering that $NiTe_2$ lacks nuclear spin [22,23] as well as obvious magnetic interaction [Figs. S4 in SM] [24], this feature cannot be ascribed to a nuclear Schottky anomaly or spin fluctuations. Therefore, the anomaly in $C_p(T)/T$ might originate from magnetic impurities, $i.e.$ trace amounts of excess Ni.

From $C_p(T)$, we can extract the phenomenological parameters of both the normal and superconducting states. For the normal state above $T_c$, the $C_p(T)$ data from 0.7 K to 3 K can be well fitted using the formula $C_p = \gamma_n T + \beta_3 T^3 + \beta_5 T^5$ [inset in Fig. 3(c)]. The fitted parameters are the Sommerfeld coefficient $\gamma_n$ = 6.43 mJ/mol·$K^2$, the lattice coefficient $\beta_3$ = 0.54 mJ/mol·$K^4$, and the high-order lattice coefficient $\beta_5$ = 0.01 mJ/mol·$K^6$. We then extract the electronic contribution using $C_e(T) = C_p(T) - \beta_3 T^3 - \beta_5 T^5$ (the possible impurity contribution will also be included). The curve of $C_e(T)/T$ as a function of temperature is shown in Fig. 3(d). Under zero field, $C_e(T)/T$ is linear in the temperature range from 60 mK to 120 mK, and its extrapolation yields a negative intercept at $T = 0$. This implies the presence of a high-power term in $C_e(T)/T$ below 60 mK, which is indicative of a fully gapped superconductor [19]. Furthermore, the jump in the $C_e(T)$ at the superconducting transition ($\delta C_e/C_e$ = 1.03) is smaller than the standard BCS value of 1.43. The reduced jump intensity may be caused by either the anisotropy of a single band or by two-band superconductivity [25]. Given the observed two-band features in the phase diagram, the lower jump intensity is more likely attributable to two-band superconductivity rather than to the anisotropy of a single band. To estimate the superconducting gaps, we extract $\delta C_{sc} = [C_e(T, 0\ \text{Oe}) - C_e(T, 100\ \text{Oe})]$ to remove the contributions from magnetic defects, and fit it using a two-band $s$-wave superconducting model, where $C_{sc} = \delta C_{sc} + \gamma_n T \propto p\ \exp(-\Delta_1/k_B T) + (1-p)$

$\exp(-\Delta_2/k_B T)$ [26] (inset in Fig. 3(d)). Here, $\Delta_1$ and $\Delta_2$ are the superconducting gaps, and $p$ is the weight of the electronic density of states corresponding to $\Delta_1$. In the fitting, $T_c$ in $C_e$ is defined as 264 mK according to entropy conservation between the normal and superconducting states (Fig. S5 in the SM). The fitting yields $p$ = 0.73, $\Delta_1$ = 0.11 meV, and $\Delta_2$ = 0.03 meV. Additionally, we extract the Ginzburg-Landau (G-L) coherence length $\xi_{GL}(0)$ = 311 nm, the G-L penetration depth $\lambda_{GL}(0)$ = 386 nm, and the G-L parameter $\kappa_{GL}(0)$ = 1.24 (Table S1 in the SM), which suggest that $NiTe_2$ could be a type-II superconductor. The above analysis establishes that $NiTe_2$ could be a type-II superconductor with fully opened superconducting gaps.

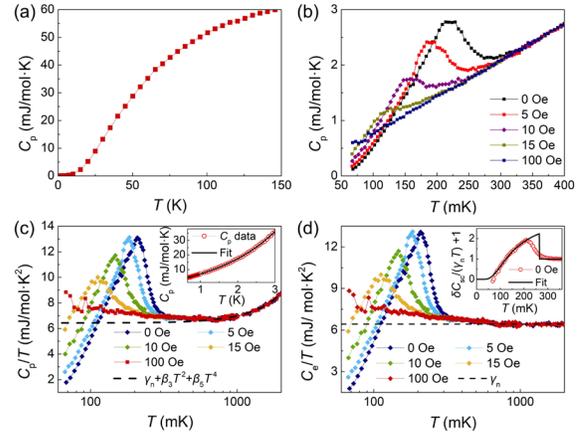

FIG. 4. Specific heat capacity of $NiTe_2$ crystals. (a) Temperature-dependent specific heat capacity, $C_p(T)$, measured from 150 K to 70 mK. (b) $C_p(T)$ below 400 mK under different $H_{\parallel c}$. (c) $C_p(T)/T$ under different $H_{\parallel c}$. (d) Temperature dependence of the electronic specific heat contribution, $C_e(T)/T$, under various $H_{\parallel c}$.

In summary, by combining STM, DFT simulations, electrical transport, and specific heat measurements, we have demonstrated the coexistence of topological surface states at $E_F$ and intrinsic superconductivity in bulk $NiTe_2$, providing a robust foundation for investigating the interplay between superconductivity and topological states. Previous studies on $NiTe_2$-based JJs have also indicated possible coupling between topological surface states (and hinge states) and superconducting currents [8,9]. The discovery of intrinsic superconductivity in $NiTe_2$ will further motivate the exploration of these coupling effects. Our findings establish a new platform for exploring the interaction between topology and superconductivity in stoichiometric transition metal dichalcogenides, and for investigating novel quantum phenomena such as topological superconductivity and finite-momentum Cooper pairing.

†Contact author: junyi_ge@t.shu.edu.cn
‡Contact author: zhe.li@hit.edu.cn

## ACKNOWLEDGMENTS

This research was supported by the National Natural Science Foundation of China (91961102, 12174242, 12174021), the Shenzhen research funding (GXWD20231130102757002), GuangDong Basic and Applied Basic Research Foundation (2514050002286). J.-Y.G. also thanks the support by the Program for Professor of Special Appointment (Eastern Scholar) at Shanghai Institutions of Higher Learning. We acknowledge the valuable discussions with Yaojia Wang and Heng Wu.

[1] M. Z. Hasan and C. L. Kane, Rev. Mod. Phys. 82, 3045 (2010).

[2] N. P. Armitage, E. J. Mele, and A. Vishwanath, Rev. Mod. Phys. 90, 015001 (2018).

[3] M. Sato and Y. Ando, Rep. Prog. Phys. 80, 076501 (2017).

[4] C. Xu et al., Chem. Mater. 30, 4823 (2018).

[5] B. Ghosh et al., Phys. Rev. B 100, 195134 (2019).

[6] S. Mukherjee et al., Sci. Rep. 10, 12957 (2020).

[7] M. Nurmamat et al., Phys. Rev. B 104, 155133 (2021).

[8] T. Le et al., Nat. Commun. 15, 2785 (2024).

[9] B. Pal et al., Nat. Phys. 18, 1228 (2022).

[10] B. S. de Lima et al., Solid State Commun. 283, 27 (2018).

[11] Z. Feng et al., Mater. Today Phys. 17, 100339 (2021).

[12] F. Zheng et al., Phys. Rev. B 101, 100505 (2020).

[13] V. D. Esin et al., Nanomaterials 12, 4114 (2022).

[14] B. T. Blue et al., Surf. Sci. 722, 122099 (2022).

[15] M. Ren et al., Phys. Rev. B 108, 235408 (2023).

[16] Y. Li et al., Proc. Natl. Acad. Sci. U.S.A. 115, 9503 (2018).

[17] A. Gurevich, Phys. Rev. B 67, 184515 (2003).

[18] W. Zheng et al., Phys. Rev. B 102, 125103 (2020).

[19] S. Kittaka et al., Phys. Rev. Lett. 112, 067002 (2014).

[20] J. S. Helton et al., Phys. Rev. Lett. 98, 107204 (2007).

[21] Y. Y. Huang et al., Phys. Rev. Lett. 127, 267202 (2021).

[22] T. Metz et al., Phys. Rev. B 100, 220504 (2019).

[23] L. Shlyk et al., J. Phys.: Condens. Matter 11, 3525 (1999).

[24] E. Uchida and H. Kondoh, J. Phys. Soc. Jpn. 11, 21 (1956).

[25] M. Zehetmayer, Supercond. Sci. Technol. 26, 043001 (2013).

[26] R. Goyal et al., J. Supercond. Novel Magn. 28, 1427 (2015).

[†]Contact author: junyi_ge@t.shu.edu.cn
[‡]Contact author: zhe.li@hit.edu.cn